\documentclass[aps,reprint,floatfix,showpacs,longbibliography]{revtex4-1}
\usepackage{graphicx}
\usepackage{amsmath}
\usepackage{amssymb}
\usepackage{amsfonts}
\usepackage{bm}        
\usepackage{dsfont}
\usepackage[mathscr]{eucal}
\usepackage[colorlinks, linkcolor=red,citecolor=blue,urlcolor=blue]{hyperref}
\usepackage[normalem]{ulem}
\usepackage[all]{hypcap}
\usepackage{multirow}
\usepackage[dvipsnames,usenames,table]{xcolor}

\newcommand{\vk}{\mathbf{k}}

\newcommand{\eq}[1]{\begin{align}#1\end{align}}

\newcommand{\nn}{\nonumber}
\newcommand{\pr}{\mathbf{r}}
\newcommand{\msr}[1]{\mathscr{#1}}
\newcommand{\mcl}[1]{\mathcal{#1}}
\newcommand{\mrm}[1]{\mathrm{#1}}

\begin{document}
\title{Disordered Chern insulator with a two step Floquet drive}

\author{Sthitadhi Roy}
\author{G. J. Sreejith}
\affiliation{Max-Planck-Institut f{\"u}r Physik komplexer Systeme, N{\"o}thnitzer Stra{\ss}e 38, 01187 Dresden, Germany}

\begin{abstract}
We explore the physics of a Chern insulator subjected to a two step Floquet drive. We analytically obtain the phase diagram and show  that the system can exhibit different topological phases characterized by presence and chirality of edge-modes in the two bulk gaps of the Floquet quasienergy spectrum, around $0$ and $\pi$. We find that the phase of the system depends on the mean but not on the amplitude of the drive. The bulk topological invariants characterizing the phases can be extracted by mapping the unitary evolution within a time period to an energetically trivial but topologically non-trivial time evolution. An extensive numerical study of the bulk topological invariants in the presence of quenched disorder reveals new transitions induced by strong disorder (i) from the different topological to trivial insulator phases and (ii) from a trivial to a topological Anderson insulator phase at intermediate disorder strengths. Careful analysis of level statistics of the quasienergy spectrum indicates a `levitation-annihilation' mechanism near these transitions.
\end{abstract}
\maketitle
\section{Introduction \label{sec:intro}}
The quantized edge response of a quantum Hall system has been understood as a reflection of a bulk topological order.\cite{TKNN1982,H1993,H1993a}
Chern insulators, a related class of systems, form a subset of what has now become an extensive area of research under the nomenclature of topological insulators.\cite{KM2005,KM2005a,BHZ2006,FKM2007,R2009,HK2010,QZ2011,book_BH2013}
They show quantum Hall-like responses in the absence of net external magnetic field but due to intrinsically broken time reversal symmetry.\cite{H88}
The associated bulk topological invariant is known as the Chern number.
Although conventionally realized in solid state systems,\cite{konig2007,zhang2009,chen2009}, recently, it has been possible to realize them in cold atomic systems in optical lattices,\cite{Hauke2012,Aidelsburger2013,Miyake2013,Jotzu2014} and photonic lattices,\cite{Rechtsman2013,Leykam2016} offering immense tunability and fine control over system parameters. 
Such realizations have generally exploited the ability of periodic modulations through light matter interactions, and rotations or mechanical deformations of confinements to effectively mimic topologically non-trivial Hamiltonians.
For instance, topological insulators have been realized by subjecting trivial insulators to a periodic drive.
\cite{OA2009,IT2010,lindner2011floquet,GFAA2011,KOBFD2011,LBDRG2013,cayssol2013floquet,DGP2013,KP2013,UPFB2014,DOM2015,d2015dynamical,PhysRevB.93.184306}
More recently, there have been extensive efforts towards classifying different topological phases of periodically driven systems\cite{KBRD2010,nathan2015topological,KS2016a, *KS2016b,DN2016,potter2016topological,roy2016abelian,*roy2016periodic,PhysRevB.93.115429}
Attempts at addressing the effect of interactions have lead to a description of their steady-state behavior in analogy to equilibrium thermodynamics\cite{Lazarides2014, *Lazarides2014a, *Khemani2016}.

Generally, periodically driven systems are described via Hamiltonian parameters varying sinusoidally in time.
Many of their qualitative aspects can however be modeled using simpler, tractable two step modulations and their $n$-step generalizations.\cite{Goldman2014}
 Such periodic two-step modulation in one dimensional systems has been shown to result in topologically protected Floquet edge modes \cite{KRBD2010,A2012,kitagawa2012observation,AO2013,KS2013,PhysRevB.87.201109,Thakurathi2013,ATD2014} and generalized to interacting\cite{Sreejith2016} and disordered\cite{gannot2015} models.

In two dimensions these systems can carry stroboscopic, chiral propagating modes localized on the edges.
Bulk-edge correspondence suggests that presence of such edge modes is associated with some bulk topological order.
Contrary to static non-interacting Chern insulators, Chern numbers are insufficient to classify the topological phases in Floquet systems due to the periodicity in the Floquet quasienergy which plays the role analogous to the energy of a static system.
For instance, Floquet systems can have trivial bulk quasienergy bands with zero Chern number coexisting with topologically protected edge states.\cite{KBRD2010,RLBL2013}
Hence, more general winding numbers that fully characterize the time evolution over one period of the bulk have been constructed\cite{KBRD2010,RLBL2013} and generalized for disordered systems.\cite{Titum2016}

In the present work, we consider a simple two-band Chern insulator model parametrized by a hopping strength and mass subjected to two-step periodic drive.
We seek to answer the question, what different topological phases can the resulting Floquet system host, both in the presence and absence of uncorrelated quenched disorder?
To address this, we first map out the exact phase diagram of the translation invariant Floquet system by locating the critical points by studying topological gap-closings in the Floquet quasienergy spectrum and characterizing the phases via winding invariants appropriate for the Floquet systems.
We find that the Floquet system has a richer set of phases than the static system - with topological phases and associated chiral edge modes  exclusive to Floquet systems. 
We also find that topological phases can appear in the Floquet systems at parameter regimes far away from those which host topological order in the static system.
Interestingly, the topological phase depends only on the mean of the two-step periodic drive and not the amplitude.
However, the amplitude does affect the gaps in the Floquet quasienergy spectrum which in turn affect the localization lengths of the edge modes.
In fact, there are regimes in the phase diagram where there exist non-topological gap closings, which do not lead to any phase transition but cause the Floquet edge modes to disappear through a divergence in their localization length.

\begin{figure}
\includegraphics[width=\columnwidth]{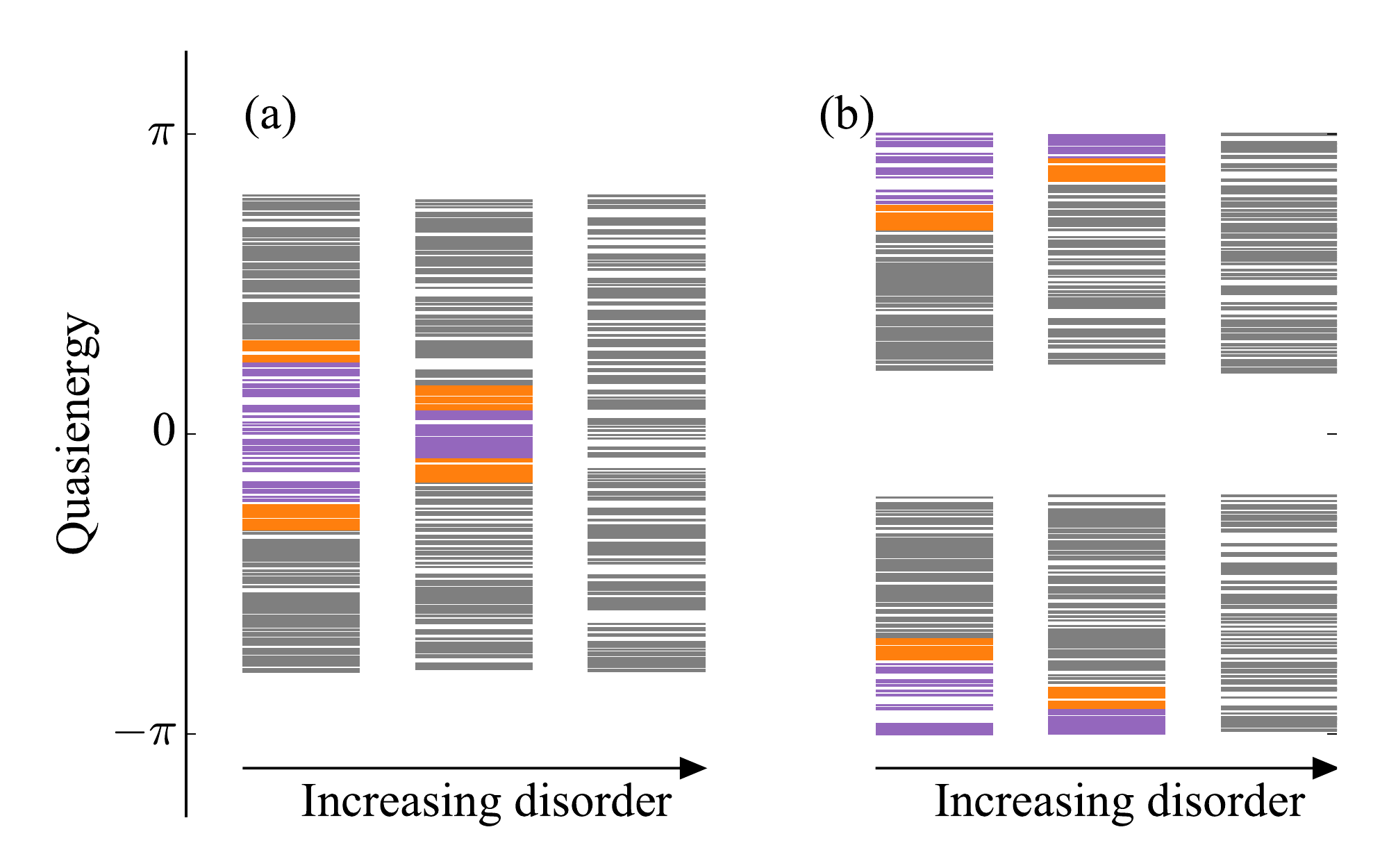}
\caption{The levitation and annihilation mechanism for Floquet systems is shown schematically, for the case of edge modes in the (a) $0$-gap and the (b) $\pi$-gap. The bulk localized, the bulk delocalized and the edge modes in the spectrum are depicted gray, orange and purple colors respectively.}
\label{fig:la}
\end{figure}

We then turn towards the effect of disorder on the phase diagram.
The interplay of disorder-induced localization\cite{Anderson1958} and topological order, in both one and two dimensions has formed a significant area of research.\cite{prodan2011disordered,loring2011disordered,MHSP2014,SP2014,ABK2015,NKR2007,RMOF2007,EM2007,OAN2007,OFRM2008,PHB2010,EV2015,CLM2015,CGLM2016}
Although in a static two-dimensional system, disorder localizes all bulk states, systems with quantum Hall-like topological order necessarily have a narrow window of energy possessing delocalized states, which can be further argued from the response of the system to gauge flux insertion.\cite{halperin1982}
The Floquet system considered in this work also has a similar behavior, however quasienergies of the delocalized states depend on the  particular underlying topological phase.
We characterize the phases in the presence of disorder by calculating the appropriate Floquet invariants generalized to include disorder.\cite{KBRD2010,Titum2016}
Analysis of the bulk order indicates a transition to trivial phase at strong disorder. Robustness of the Floquet topological phases and hence the critical disorder for a disorder-induced topological transition is intimately connected to the localization lengths of the disorder free edge states and hence the gaps in the Floquet quasienergy spectrum.
We find that the disorder-induced topological phase transitions take place via a ``levitation and annihilation''\cite{OAN2007} mechanism generalized for Floquet systems.
As shown schematically in Fig.~\ref{fig:la}, the delocalized states are present close to the edges of the bulk bands immediately surrounding the edge modes.
As disorder is increased, the delocalized states from two bands levitate towards the gap before finally meeting and annihilating each other at the critical disorder.
We present evidence for this mechanism by carefully analyzing energy resolved level-spacing statistics for the Floquet quasienergies.
In the case of of multiple edge modes with corresponding sets delocalized bulk states, we find that levitation and annihilation always takes place between states from different bands.

This is qualitatively different from the anomalous Floquet-Anderson insulator discussed in Ref.[\onlinecite{Titum2016}], where Floquet bulk bands with zero Chern number do not posses any delocalized state and hence the edge states persist at all quasienergies.
On the contrary the Floquet bulk bands in the model studied in this work always posses a finite Chern number in a topological phase necessitating the presence of delocalized bulk states.

The rest of the paper is organized as follows. We start with reviewing the topological invariants for the Floquet system in Sec.~\ref{sec:invariants}. Sec.~\ref{sec:model} describes the Chern insulator model and the two-step Floquet drive. We present a detailed analysis of the phase diagram in the absence of disorder in Sec.~\ref{sec:phasediag_clean}. The effect of disorder on the phase diagram and the analysis of level spacing statistics is discussed in Sec.~\ref{sec:phasediag_dis}. Finally, we summarize the main results of the paper and provide future outlook in Sec.~\ref{sec:discussions}.

\section{Topological Invariants for the Floquet System}
\label{sec:invariants}
Chern numbers of single particle bands provide a complete characterization of the edge modes in a static system in the absence of any symmetries.\cite{ryu2010topological}
Chern number of a band equals the difference between the chirality of edge modes above and below the band. Since the spectrum of the Hamiltonian is bounded, there are $0$ chiral modes below the lowest energy band and above the highest band. 
As a result, the Chern number of the bands completely determine the counting of edge modes (Chirality determines the number of modes, in the absence of any symmetries).

In close analogy to the notion of energy spectrum of a static Hamiltonian, one can define the quasienergy spectrum for a Floquet system. Quasienergies correspond to the argument of the complex unimodular eigenvalues of the unitary time evolution over a period of time. The quasienergies are periodic and are well defined modulo the frequency of the drive {\it i.e}  $\omega\equiv\omega+\frac{2\pi}{T}$. The quasienergy spectrum also has bands analogous the energy bands of a static Hamiltonian. Chern number of a such a Floquet band is again equal to the difference between the chirality of modes above and below. However, periodicity of the quasienergies implies that there is no notion of highest or lowest bands near which the number of edge modes can be fixed. The result is that, Chern numbers do not completely characterize the number of edge modes. A striking instance of this is the anomalous Floquet Chern insulator in which all bands of the spectrum have zero Chern number but carry a chiral mode in every gap between the bands.\cite{RLBL2013}

A bulk invariant which correctly characterizes such a Floquet system was introduced in Ref~[\onlinecite{RLBL2013}] and generalized to the case of disordered systems in Ref~[\onlinecite{Titum2016}]. We use these invariants extensively in this work to numerically characterize the phases exhibited by our model system. We present here a brief intuitive explanation and motivation for this invariant. The main result of the discussion is contained in Eq.~\eqref{DisorderFreeInvariant}.

Let $U(t)$, $t\in [0,T]$ be the time evolution of our system. For simplicity, in this section, we shall assume units where $T=1$. Consider a system with a modified time evolution of the following form
\begin{equation}
\mathcal{U}\left(t\right)=\begin{cases}
U\left(2t\right) & t\in\left[0,\frac{1}{2}\right]\\
\exp\left[-\imath 2H_{{\rm eff}}\left(1-t\right)\right] & t\in\left[\frac{1}{2},1\right]
\end{cases},
\label{eq:returnmap}
\end{equation}
where $H_{\rm eff}$ is the effective Hamiltonian, defined as
\begin{equation}
\exp\left[-\imath H_{{\rm eff}}\right]=U\left(1\right).
\label{eq:heff1}
\end{equation}

\begin{figure}
\includegraphics[width=\columnwidth]{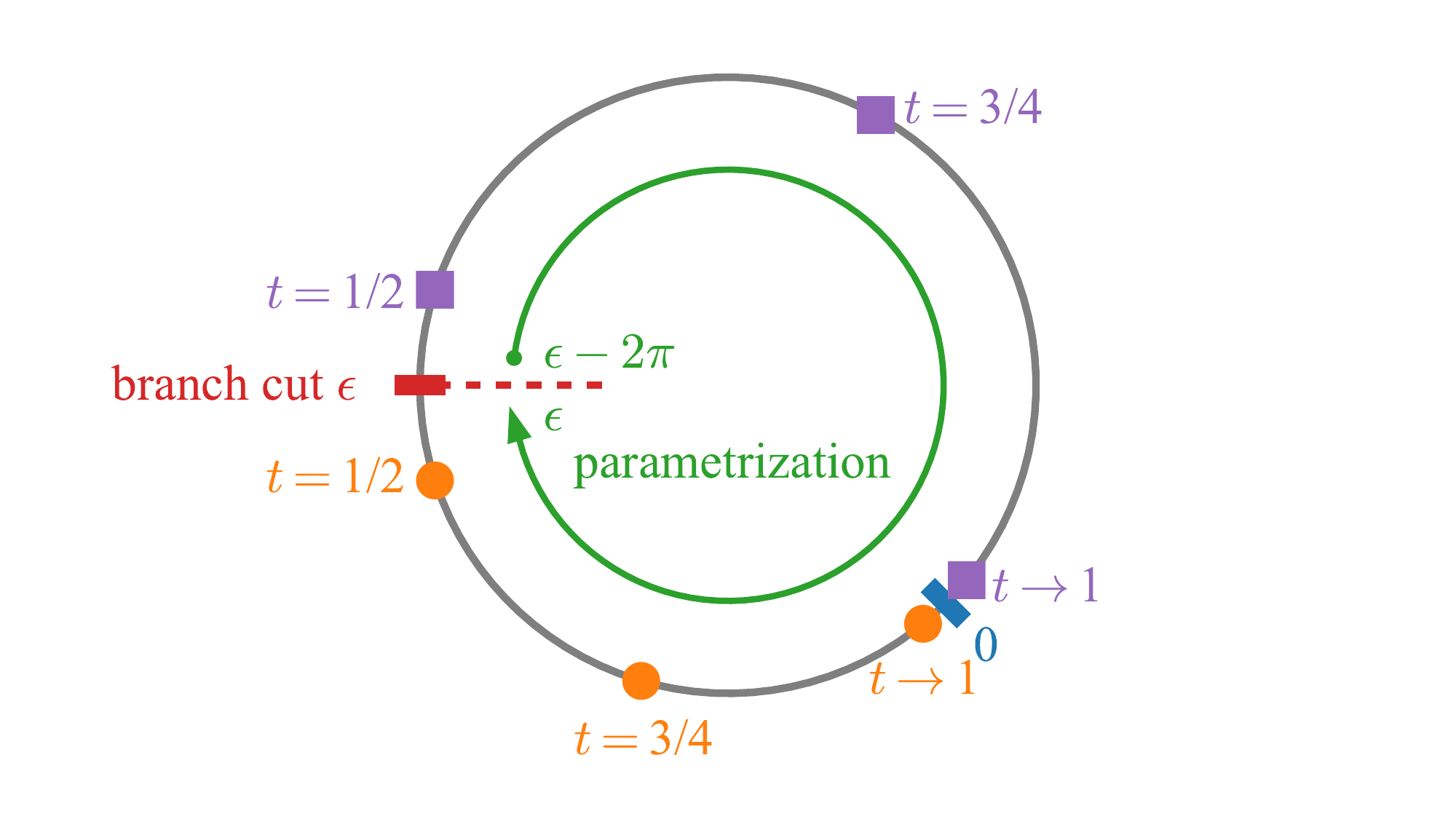}
\caption{The quasienergies of the unitary operator $\mathcal{U}(t)$ can be represented on a circle parametrized by $[\epsilon-2\pi,\epsilon]$. The branch cut, $\epsilon$, and 0 define two disjoint parts of this circle.
Eigenvectors of $\mathcal{U}(t)$ do not change between $t=1/2$ and $t=1$ but the eigenvalues change.  Depending on the location of an eigenvalue at $t=1/2$, its value at $t>1/2$ drifts towards $0$ along two opposite directions. The circle and the square represent the evolution of the two eigenvalues as a function of $t$.}
\label{Fig:BcShownOnACircle}
\end{figure}

Eq.~\eqref{eq:heff1} does not uniquely define $H_{\rm eff}$, and as will be discussed below, the definition of the winding number makes use of this freedom to probe the edge modes in different band gaps. If the eigenvalues and vectors of $U(1)$ are $\{u_i\}$ and $\{\left \vert i \right \rangle\}$, for some choice $\epsilon\in[0,2\pi]$ of the branch cut, we can define $H_{\rm eff}^\epsilon$ as 
\begin{eqnarray}
H_{\rm eff}^\epsilon = \sum_{i} \omega_i \left\vert i \right \rangle \left \langle i \right \vert \text{, } \omega_i = -\arg^{\epsilon}(u_i)
\end{eqnarray}
where $\arg^x$ is defined to be between $x-2\pi$ and $x$.
With this choice, the modified unitary operator $\mathcal{U}(t)$, for the interval $t\in [\frac{1}{2},1]$ takes the form:
\begin{equation}
\mathcal{U}(t) = \sum_{i} \exp\left[ -\imath \omega_i 2(1-t)\right] \left\vert i \right \rangle \left \langle i \right \vert
\end{equation}
As $t$ goes from $\frac{1}{2}$ to $1$, $\mcl{U}(t)$ interpolates from $U(1)$ to $\mathds{I}$. The eigenvectors remain the same but the eigenvalues change from $e^{-\imath \omega_i}$ to $1$. During the interpolation, the eigenvalues on the two sides of the branch cut drift towards $1$ along two different paths as shown in Fig.~\ref{Fig:BcShownOnACircle}.

\begin{figure}
\includegraphics[width=\columnwidth]{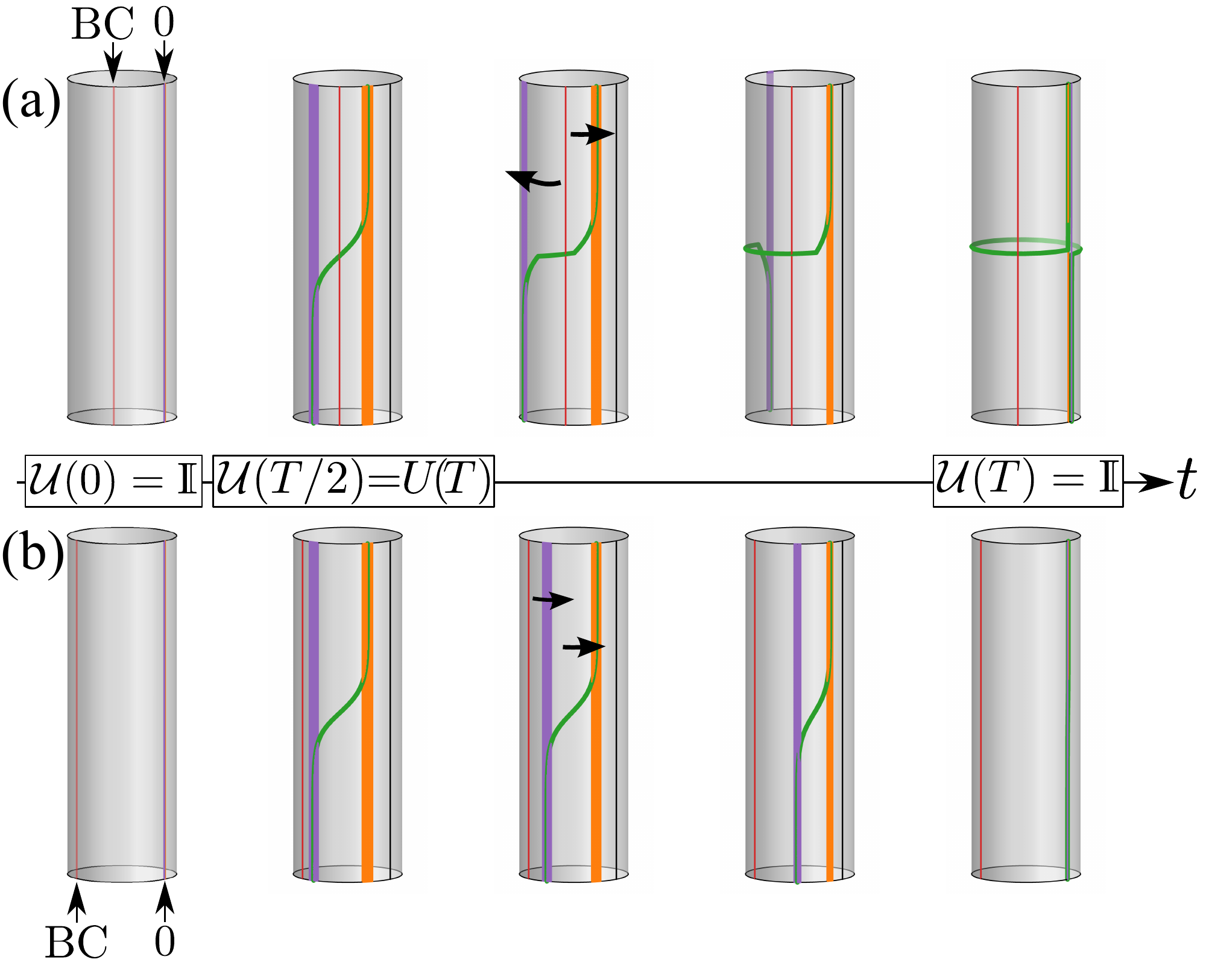}
\caption{Eigenvalues of the unitary operator $\mathcal{U}$ of a system with semi-periodic boundary conditions represented on a cylinder, with the momentum quantum number along the periodic direction represented by the length of the cylinder and the quasienergies represented along the circumference. The panels (a) and (b) show the evolution of the eigenvalues of $\mathcal{U}(t)$. All eigenvalues are $1$ at $t=0$ (left panels), spectrum of the physical unitary operator $U=\mathcal{U}(1/2)$ is shown in the second panel from the left. Evolution of eigenvalues from $t=1/2$ to $1$ can happen in two different ways as shown in the (a) and (b) depending on the choice of the branch cut.}
\label{Fig1}
\end{figure}

Consider the operator $\mcl{U}(t)$ \eqref{eq:returnmap} for a two dimensional Floquet system with periodic boundary conditions along the $y$-direction such that the momentum $k_y$ is a good quantum number, and open boundaries along the $x$-direction with chiral edge modes whose quasienergies are within the bulk-gap surrounding some quasienergy $\omega_{\mrm{edge}}$. This can be represented on a cylinder as shown in Fig.~\ref{Fig1}(a) (second panel), where the length of the cylinder represents the $k_y$-axis and the circular direction represents the quasienergies. If the branch cut $\epsilon$ is chosen to lie in the same gap that contains $\omega_{\mrm{edge}}$, the operator $\mathcal{U}(t)$ for $t\in[1/2,1]$ has an edge mode in the same gap. As $t$ goes from $1/2$ to $1$, this edge mode is stretched in such a way that $\mathcal{U}(t\to 1)$ has an edge mode that winds around the entire cylinder, whereas all the bulk modes shrink to quasienergy $0$. This is schematically shown in Fig.~\ref{Fig1}(a). The contrary scenario where the branch cut is chosen to be in a different gap results in the edge modes also shrinking to $0$ as shown in Fig.~\ref{Fig1}(b).

The total number of chiral edge modes of $U(1)$ at quasienergy $\epsilon$ is the same as that of $\mathcal{U}^\epsilon(t\to 1)$ where $\mcl{U}^\epsilon$ is the modified time evolution operator \eqref{eq:returnmap} defined with the branch cut placed at $\epsilon$.
This is captured by the winding number $\int  \frac{dk_y}{2\pi} ~\mathrm{Tr} [\mathcal{U}^{\epsilon\dagger} \partial_{k_y} \mathcal{U}^\epsilon]$, since this is identical to the total winding of the quasienergies $\sum_i \int \frac{dk_y}{2\pi\imath} \partial_{k_y} \omega_i $. The number of chiral modes on a single edge is obtained by projecting the integrand to sites on one half of the system as
\begin{multline}
n(\epsilon)=\int  \frac{dk_y}{2\pi\imath} \mathrm{Tr}\, [\msr{P} \mathcal{U}^{\epsilon\dagger} (1) \partial_{k_y} \mathcal{U}^\epsilon(1)] \\ =\int_0^1 dt \, \partial_{t} \int  \frac{dk_y}{2\pi\imath} \mathrm{Tr}\, [\msr{P}  \mathcal{U}^{\epsilon\dagger} (t) \partial_{k_y} \mathcal{U}^\epsilon(t)], 
\end{multline}
where $\msr{P}$ is the projector onto one half of the system, $\mathrm{Tr}$ represents the trace over all sites, and $n(\epsilon)$ is the number of chiral edge modes at quasienergy $\epsilon$. The second equality arises from the fact that the argument of $\partial_t$ is a real valued non-singular quantity that changes from $0$ to $n$ as $t$ changes from $0$ to $1$. Note that, the integrand in the above winding number is defined for an open system. The support of the integrand can be moved to the bulk by adding a total derivative $ \partial_{k_y}  \mathrm{Tr}\, [\msr{P}  (\mathcal{U}^{\epsilon\dagger} \partial_t  \mathcal{U}^\epsilon)]$ to the integrand, giving
\begin{equation}
n(\epsilon) = \int \frac{dk_ydt}{2\pi\imath} \mrm{Tr}\big[\mathcal{U}^{\epsilon\dagger} \partial_{k_y} \mathcal{U}^\epsilon [\msr{P},\mathcal{U} ^{\epsilon\dagger} \partial_{t} \mathcal{U}^\epsilon]\big].
\end{equation}
Since $\msr{P}$ is identity close to the edges, the commutator in the integrand is non-zero only in the bulk, around the region where the diagonal of $\msr{P}$ changes from $1$ to $0$. Assuming $\mathcal{U}$ has a finite range, it can be replaced with the unitary operator $\mathcal{U}$ defined for a system with periodic boundary conditions. Expressing the operators in the momentum basis, we arrive at 
\begin{equation}
n(\epsilon) = \int \frac{d^2\vk dt}{8\pi^2}\mathrm{Tr}\; \big[\mathcal{U}^{\epsilon\dagger} \partial_{t} \mathcal{U}^\epsilon \left[\mathcal{U}^{\epsilon\dagger} \partial_{k_x} \mathcal{U}^\epsilon,\mathcal{U}^{\epsilon\dagger} \partial_{k_y} \mathcal{U}^\epsilon\right]\big].
\label{DisorderFreeInvariant}
\end{equation}
$\mcl{U}^\epsilon\equiv\mathcal{U}^\epsilon(t,k_x,k_y)$ is the modified time evolution defined for a system with periodic boundary conditions,
\begin{equation}
\mathcal{U}^\epsilon\left(t,k_x,k_y\right)=\begin{cases}
U_\vk\left(2t\right) & t\in\left[0,\frac{1}{2}\right]\\
\exp\left[-\imath 2H^\epsilon_{{\rm eff},\vk}\left(1-t\right)\right] & t\in\left[\frac{1}{2},1\right]
\end{cases},
\end{equation}
where $H^\epsilon_{{\rm eff},\vk}$ is the effective Hamiltonian defined with all quasienergies inside the interval $[\epsilon-2\pi,\epsilon]$. 

Note that the original time evolution $U$ does not return to itself at $t=T$ since $U(t=T)\neq U(t=0)$. The modified time evolution, while preserving the edge counting (in the gap selected by the branch cut) satisfies $\mathcal{U}(t=T)= \mathcal{U}(t=0)$. This allows us to compactify time and characterize $\mathcal{U}$ using homotopy classes of maps from $S^1_t\times S^1_{k_x}\times S^1_{k_y}$ to the unitary group.\cite{bott1978} The characterizing invariant is given by Eq.~\eqref{DisorderFreeInvariant}.

The Floquet systems that we consider in our study has two gaps - around $0$ and $\pi$, and the topology of the drive is defined by the counting of the edge modes in the two gaps. Thus the phases of the Floquet system can be fully characterized by the pair $(\nu_0,\nu_\pi)$ 
\begin{equation}
\nu_0=n(0)\text{ and }\nu_\pi=n(\pi).
\end{equation}
Although insufficient to characterize the phases of the Floquet system, the Chern number of the Floquet bulk bands, $\mcl{C}$, within a quasienergy window $\omega\in[0,\pi]$ is the difference in the number of chiral edge modes at $0$ and $\pi$ and can be formally expressed as
\eq{\mcl{C} = & \nu_0-\nu_\pi \nn\\
=&\int\frac{d\mathbf{k}}{4\pi}\mrm{Tr}\big[\msr{P}_\vk[\partial_{k_x}\msr{P}_\vk,\partial_{k_y}\msr{P}_\vk]\big]\label{eq:cn},
}
where $\msr{P}_\vk$ is the projector onto eigenstates of $H_{\mrm{eff},\vk}^\epsilon$ having quasienrgies in window $[0,\pi]$.

The winding number described above can be generalized to the case of disordered systems by considering a periodic superlattice constructed with the entire disordered system as the unit cell. Presence of an edge mode in a single disordered sample implies the presence of an edge mode also in the superlattice. The above expressions can now be used to probe the presence of an edge mode in this system also. The quasimomenta on this lattice appear as twisted boundary conditions across each unit cell. After a change in basis, the twisted boundary conditions can be reinterpreted as flux-insertions through the two holes of the torus. This results in a form of the winding number similar to the one in Eq.~\eqref{DisorderFreeInvariant} but $\mathcal{U}(t,k_x,k_y)$ is now replaced by $\mathcal{U}(t,\theta_x,\theta_y)$ representing the unitary operator for a system with periodic boundary conditions but with fluxes $\theta_{x,y}$ through the holes of the two-torus representing the two spatial directions. We defer further details to Sec.~\ref{sec:phasediag_dis} where we describe our model in the presence of disorder.

\section{Model and two-step Floquet \label{sec:model}}

\subsection{Static properties \label{sec:static}}

To describe a two-band Chern insulator, we employ a model of spinless fermions on a square lattice \cite{HK2010,QZ2011,book_BH2013} described by the Hamiltonian
\eq{
\nn \mcl{H} = -\frac{J}{2}\sum_\pr [c^\dagger_\pr(\sigma^z-\imath\sigma^x)&c_{\pr+\hat{x}}+c^\dagger_\pr(\sigma^z-\imath\sigma^y)c_{\pr+\hat{y}}\\&+\mrm{h.c}]
+M\sum_\pr c^\dagger_\pr\sigma^zc_{\pr},
\label{eq:ham_real}
}
where the fermionic creation (annihilation) operators $c^\dagger_\pr (c_\pr)$ at site $\pr$ represent a two-spinor of operators for each sublattice, $c^\dagger_\pr = (c^\dagger_{\pr,\alpha},c^\dagger_{\pr,\beta})$.
The model is closely related to Haldane`s honeycomb model\cite{H88} in the sense that time reversal symmetry is broken via complex hoppings. Real time dynamics arising from periodic driving in similar two-level systems has been studied in Ref-\onlinecite{Mukherjee2016}.
The Hamiltonian \eqref{eq:ham_real} can be reduced to family of two-level Hamiltonians in reciprocal space, each corresponding to a momentum mode which can be represented using Pauli matrices as $\mcl{H}_\vk = \mathbf{d}_\vk\cdot\bm{\sigma}$ where, 
\eq{
	\mathbf{d}_\vk=-\{J\sin k_x,J\sin k_y,J\cos k_x+J\cos k_y-M\}.
	\label{eq:ham_k}
}
The vector $\mathbf{d}_\vk$ in Eq.~\eqref{eq:ham_k} represents a pseudospin texture in the Brillouin zone, whose skyrmion number gives the Chern numbers of the two underlying bands. Since they have to sum up to zero, they are negative of each other.
The model hosts topological phase transitions at $M=0$ and $M/J=\pm2$ with the Chern number of the lower band being 0 for $\vert M/J\vert>2$ and $\mrm{sgn}(M/J)$ for $\vert M/J\vert<2$.
The transitions are accompanied by linear gap closings in the energy spectrum at the high-symmetry points of the Brillouin zone, they being $(0,0)$ for $M=2J$, $(\pi,\pi)$ for $M=-2J$, and $(0,\pi)$ and $(\pi,0)$ for $M=0$.
Consistent with the understanding that ground state spinors of two-level systems (determined by $\mathbf{d}_\vk\cdot\bm{\sigma}$) in different topological phases are orthogonal at least at one point in the Brillouin zone, it turns out that the $\mathbf{d}_\vk$s \eqref{eq:ham_k} in different adjacent topological phases are indeed anti-parallel \footnote{The modulus of the overlap squared of ground state spinors corresponding to $\mathbf{d}_1\cdot\bm{\sigma}$ and $\mathbf{d}_2\cdot\bm{\sigma}$ is given by $(1+\mathbf{d}_1\cdot\mathbf{d}_2)/2$.} at the gap-closing high-symmetry point(s).
This can be confirmed by analyzing the skyrmion textures at the high-symmetry points which turn out to be
\eq{
	\mathbf{d}_{(0,0)} &= \{0,0,2J-M\}, \nn\\
	\mathbf{d}_{(\pi,\pi)} &= \{0,0,-2J-M\}, \label{eq:skyrmion_hsp}\\
	\mathbf{d}_{(0,\pi)} &= \mathbf{d}_{(\pi,0)}=\{0,0,M\}.\nn 
}
It is sufficient to focus only at the high-symmetry points as the skyrmion textures corresponding to two different values, $M_A$ and $M_B$, can become anti-parallel only at the high-symmetry points.
This is because, it is evident from Eq.~\eqref{eq:ham_k} that $\mathbf{d}_{A,\vk} = -\mathbf{d}_{B,\vk}$ implies $d^x_\vk=0=d^y_\vk$ as $d^x_\vk$ and $d^y_\vk$ are independent of $M$.
Hence, critical points in the parameter space can be deduced from the zeros of $\mathbf{d}_\vk$ at the high-symmetry points in Eq.~\eqref{eq:skyrmion_hsp}.

\subsection{Two-step Floquet \label{sec:twostep}}
We subject the Chern insulator model \eqref{eq:ham_real} to a two-step Floquet drive by periodically modulating the mass-term in the Hamiltonian as
\eq{
	M(t) = \begin{cases}
M_A; ~~ nT<t<(n+1/2)T\\
M_B; ~~ (n+1/2)T<t< (n+1)T.
\end{cases}
\label{eq:Mt}
}
The resulting time-periodic Hamiltonian is denoted as
\eq{
	\mathcal{H}_\vk(t) = \begin{cases}
\mathbf{d}_{A,\vk}\cdot\bm{\sigma}; ~~ nT<t<(n+1/2)T\\
\mathbf{d}_{B,\vk}\cdot\bm{\sigma}; ~~ (n+1/2)T<t< (n+1)T,
\end{cases}
\label{eq:ht_k}
}
where
\eq{
	\mathbf{d}_{X,\vk}=-J\{\sin k_x,\sin k_y,\cos k_x+\cos k_y-\frac{M_{X}}{J}\}.
	\label{eq:ham_mom1}
}
In the rest of the paper, we set $T=1$ and work in a parameter space spanned by $J$, $M_A$, and $M_B$.
The properties of a periodically driven system are governed by the time-evolution operator over one period, $U(1)=U_\mcl{F}$, often dubbed as the Floquet operator.
For the time-periodic Hamiltonian \eqref{eq:ht_k}, $U_{\mcl{F},\vk}$ can be expressed as 
\eq{
	U_{\mcl{F},\vk} &= e^{-\imath\mathbf{d}_{B,\vk}\cdot\bm{\sigma}/2}e^{-\imath\mathbf{d}_{A,\vk}\cdot\bm{\sigma}/2}\nn\\
	&=d_{0,\vk}\mathds{I}_2 - \imath~\mathbf{d}_{\mrm{eff},\vk}\cdot\bm{\sigma},
	\label{eq:floq}
}
and $\omega_{\pm,\vk}=\pm\cos^{-1}d_{0,\vk}$ are the Floquet quasienergies which also satisfy $\omega_{\pm,\vk}=\pm\sin^{-1}\vert \mathbf{d}_{\mrm{eff},\vk}\vert$.
For our model, $\mathbf{d}_{\mrm{eff},\vk}$ can be explicitly obtained by using Eq.~\eqref{eq:ham_mom1} in Eq.~\eqref{eq:floq} which gives
\eq{
\nn	\mathbf{d}_{\mrm{eff},\vk}=&\sin(d_{A,\vk}/2)\cos(d_{B,\vk}/2)\hat{\mathbf{d}}_{A,\vk}+\\
\nn &\cos(d_{A,\vk}/2)\sin(d_{B,\vk}/2)\hat{\mathbf{d}}_{B,\vk}+\\
 &\sin(d_{A,\vk}/2)\sin(d_{B,\vk}/2)\hat{\mathbf{d}}_{A,\vk}\wedge\hat{\mathbf{d}}_{B,\vk}.
 \label{eq:d_eff}
}
\section{Phase diagram for translation invariant system \label{sec:phasediag_clean}}
The topological phase transitions in the Floquet system can be understood by examining the nature of the gap closings in the spectrum of the Floquet quasienergies.
The gapless points in the quasienergy spectrum can be found by setting $\mathbf{d}_{\mrm{eff},\vk}=0$.
This naturally implies that, at the gapless points, $d_{0,\vk}=1$ or $d_{0,\vk}=-1$.
The former corresponds to the gap closing at $\omega_\pm=0~[\mrm{mod} 2\pi ]$where as the latter corresponds to the ones at $\omega_\pm =\pm\pi~[\mrm{mod} 2\pi ]$.

We find that, if the gap-closings happen at the high-symmetry points in the Brillouin zone, it leads to a topological phase transition accompanied by a change in $\nu_{0}$ or $\nu_{\pi}$, the former corresponding to a gap-closing at $\omega=0$ where as the latter to $\omega=\pi$. 
Interestingly, we also find that the Floquet drive can lead to gap-closings in the quasienergy spectrum elsewhere in the Brillouin zone which do not correspond to any topological transition, but lead to disappearance of the Floquet topological edge modes at these singular points (lines) in the phase diagram.
This has further ramifications on the robustness of these edge states and the quantitative nature of the phase diagram in the presence of disorder discussed in Sec.~\ref{sec:phasediag_dis}.

Closing of a gap in the Floquet quasienergy spectrum necessitates $\hat{\mathbf{d}}_{A,\vk}\wedge\hat{\mathbf{d}}_{B,\vk}$ itself to be zero, or the coefficient of $\hat{\mathbf{d}}_{A,\vk}\wedge\hat{\mathbf{d}}_{B,\vk}$ in Eq.~\eqref{eq:d_eff} to be zero, as $\hat{\mathbf{d}}_A\wedge\hat{\mathbf{d}}_B$ is perpendicular to both $\hat{\mathbf{d}}_A$ and $\hat{\mathbf{d}}_B$.
As argued in Sec.~\ref{sec:static}, the former can happen only at the high-symmetry points in the Brillouin zone.
Inspection of Eq.~\ref{eq:d_eff} reveals that, at these high-symmetry points, $d_\mrm{eff}^{x}=0=d_\mrm{eff}^{y}$.
Hence, by tuning the parameters of the Floquet system, one can flip the sign of $d_\mrm{eff}^{z}$ effectively making the skyrmion texture anti-parallel at the high-symmetry points leading to a phase transition. 
We exhaustively study these topological transitions in subsection~\ref{sec:tqpt}.

Careful analysis of Eq.~\ref{eq:d_eff} also shows that, for the latter case, where the coefficient of $\hat{\mathbf{d}}_{A,\vk}\wedge\hat{\mathbf{d}}_{B,\vk}$ vanishes, for the vector $\mathbf{d}_\mrm{eff}$ to vanish, $d_A = 2n\pi$ and $d_B=2m\pi$ necessarily. 
We find that this can happen generically anywhere in the Brillouin zone away from the high-symmetry points and they do not correspond to any phase transitions. 
We discuss these gap closings in subsection~\ref{sec:anomalous_gapclosing}.

\subsection{Topological phase transitions \label{sec:tqpt}}

\begin{figure}
\includegraphics[width=\columnwidth]{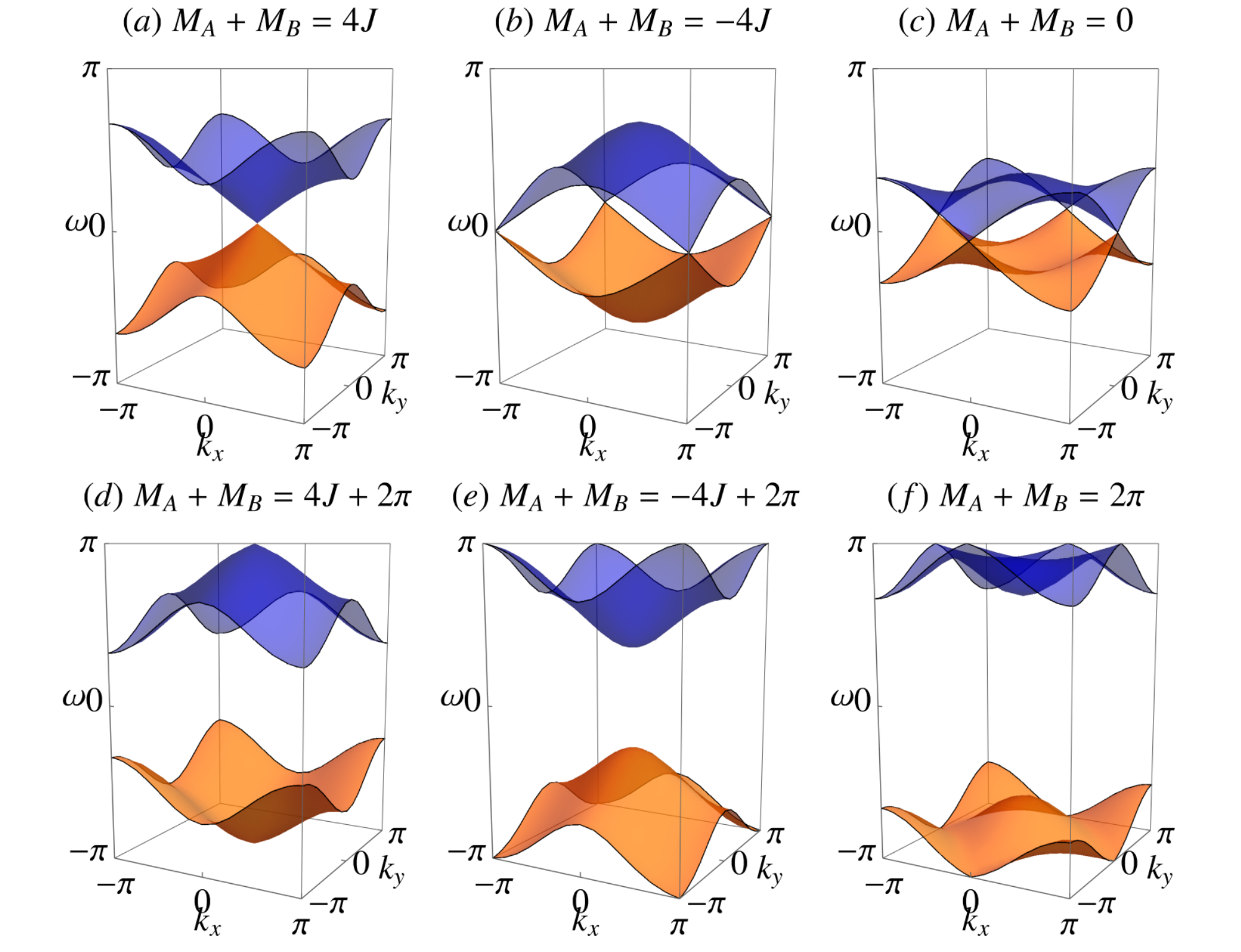}
\caption{The Floquet quasienergy spectrum $\omega_{\pm,\vk}$ at different critical points. (a)-(c) show the spectrum at the critical points which correspond to gap-closings at $\omega=0$ while (d)-(f) correspond to that at $\omega=\pm\pi$. For the plots, $J=\pi/6$ and $M_A=1$.}
\label{fig:qen_crit}
\end{figure}

As argued above, for a topological transition to occur in the Floquet system, $\hat{\mathbf{d}}_{A,\vk}\wedge\hat{\mathbf{d}}_{B,\vk}=0$, implying $\hat{\mathbf{d}}_{A,\vk}$ and $\hat{\mathbf{d}}_{B,\vk}$ are mutually parallel or anti-parallel, which can happen only at the high-symmetry points in the Brillouin zone as argued at the end of Sec.~\ref{sec:model}.
At these points, the Floquet quasienergies have the forms
\eq{
\omega_{\pm,(0,0)} &= \pm(4J-M_A-M_B)/2,  \nn\\
\omega_{\pm,(\pi,\pi)} &= \pm(4J+M_A+M_B)/2, \label{eq:qen_hsp}\\
\omega_{\pm,(0,\pi)} &=\omega_{\pm,(\pi,0)}= \pm(M_A+M_B)/2. \nn
}
By setting of $\omega_{\pm,\vk}=0$ and $\omega_{\pm,\vk}=\pi$ in Eq.~\eqref{eq:qen_hsp}, the critical points can be obtained as
\eq{
M_A+M_B = 4J\eta + (4n+2\zeta)\pi,
\label{eq:criticalpoints}
}
where $\zeta$ takes values 0(+1) for a transition with a gap closing at $\omega=0(\pi)$ and $\eta$ takes values -1, 0 and +1 depending on the particular symmetry point at which the gap closes.
One of the most important observations from the expression for the critical points \eqref{eq:criticalpoints} is, they only depend on $M_A+M_B$, which physically means that the topological properties of the Floquet system depend only on the mean of the binary drive and not on the amplitude.
Moreover, Eq.~\eqref{eq:criticalpoints} also reveals that the phase diagram is periodic in $M_A+M_B$ with a period of $4\pi$.
Representative Floquet quasienergy spectrum at different critical points are plotted in Fig.~\ref{fig:qen_crit} showing the gap-closings at the corresponding high-symmetry points.
%

%
Having established the critical points of the Floquet system, we now characterize the transitions in terms of the change in the Floquet winding numbers $\nu_{0/\pi}$  and the Chern number $\mcl{C}$.
If there exists a topological gap-closing, then the effective Hamiltonian in the vicinity of the gapless mode has the form $\sum_{i,j=x,y}\kappa_i\msr{A}_{ij}\sigma^j + \lambda\sigma^z$, where $\lambda$ is the effective mass which goes to zero at the transition. 
Across a transition, the change in the Chern number of the band arises from the vicinity of the gapless mode.
The change is given by~\cite{book_BH2013} $\sum_\mu\mrm{sgn}(\mrm{Det[\msr{A}^\mu]})(\mrm{sgn}\lambda_+^\mu-\mrm{sgn}\lambda_-^\mu)/2$ where $\mu$ indexes the gapless momenta and $\lambda^\mu_\pm$ is the effective mass on either side of the critical point. 
Hence, the nature of transitions can be understood by studying $\mathbf{d}_{\mrm{eff},\vk}$ in the vicinity of the high-symmetry points.
We state the main results here and the details of the effective Hamiltonian are relegated to Appendix.~\ref{sec:d_eff}

In the vicinity of the gapless mode at $\vk=(0,0)$, $\mathbf{d}_{\mrm{eff,\vk}}$ is such that $\mrm{sgn(Det}[\msr{A}])=1$ and the effective mass is $\sin((M_A+M_B-4J)/2)$. 
Hence across the phase transition between two points in the parameter space such that $M_A+M_B-4J<4n\pi$ and $M_A+M_B-4J>4n\pi$, the effective mass changes from negative to positive, hence the Chern number changes by $+1$. 
Since, the topological transition is accompanied by a gap closing at $\omega=0$, $\nu_0$ also changes by +1.
Correspondingly, between two points such that $M_A+M_B-4J<(4n+2)\pi$ and $M_A+M_B-4J>(4n+2)\pi$, the Chern number changes by $-1$ consistent with the change of the sign of effective mass and consequently $\nu_\pi$ changes by $+1$.

A similar analysis at $\vk=(\pi,\pi)$ yields $\mrm{sgn(Det}[\msr{A}])=1$ and the effective mass of the form $\sin((M_A+M_B+4J)/2)$. Hence similar to the $\vk=(0,0)$ case, the Chern number changes by $+1$ across a transition at $M_A+M_B=-4J+4n\pi$ and by $-1$ across a transition at $M_A+M_B=-4J+(4n+2)\pi$ 

At both gapless modes at $\vk=(0,\pi)$ and $\vk=(\pi,0)$, the effective mass has a form $\sin((M_A+M_B)T/2)$.
However, around these gapless modes it turns out that the effective Hamiltonian has $\mrm{sgn(Det}[\msr{A}])=-1$, hence the gap closings at $\omega=0$ lead to change in the Chern number of $-2$ and those at $\omega=\pi$ lead to change in the Chern number by $+2$ as there are two inequivalent points in the Brillouin zone where the gap closes.

These set of rules completely characterize the phase diagram and the topological phase transitions of the periodically driven Chern insulator and are summarized in Table.~\ref{ta:pd}
\begin{table}
\begin{tabular}{|l|c|c|c|}
\hline
Critical points & $\Delta \mcl{C}$ & $\Delta \nu_0$ & $\Delta \nu_\pi$\\
\hline
$M_A+M_B=4J+4n\pi$ & $-1$ & $-1$ & $0$\\
$M_A+M_B=4J+(4n+2)\pi$ & $+1$ & $0$ & $-1$\\
$M_A+M_B=4n\pi$ & $+2$ & $+2$ & $0$\\
$M_A+M_B=(4n+2)\pi$ & $-2$ & $0$ & $+2$\\
$M_A+M_B=-4J+4n\pi$ & $-1$ & $-1$ & $0$\\
$M_A+M_B=-4J+(4n+2)\pi$ & $+1$ & $0$ & $-1$\\
\hline
\end{tabular}
\caption{Summary of the phase transitions of the periodically driven Chern insulator. The four columns respectively show the critical points, the changes in $\mcl{C}$, $\nu_0$, and $\nu_\pi$.}
\label{ta:pd}
\end{table}

\begin{figure}
\includegraphics[width=\columnwidth]{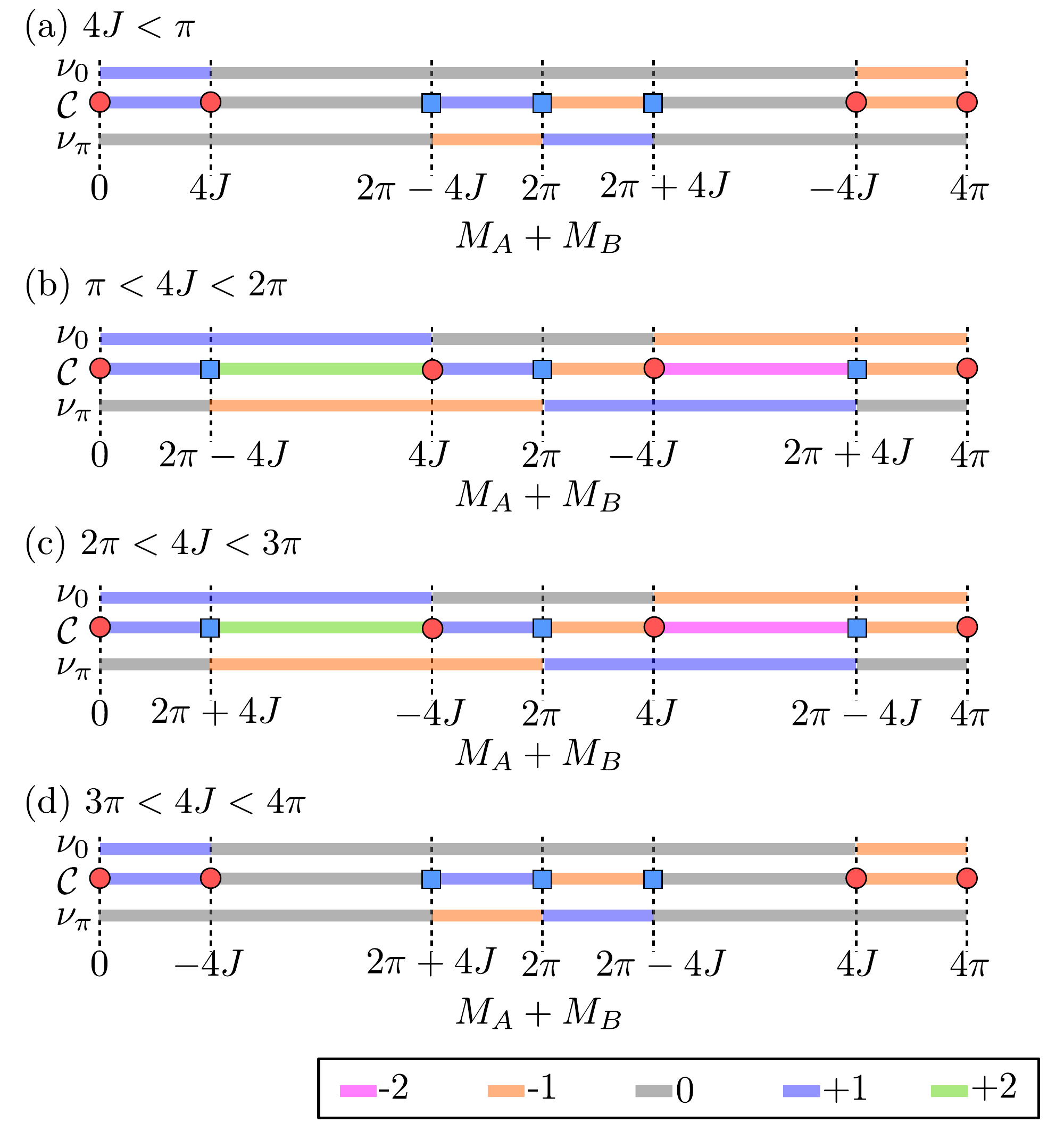}
\caption{Graphical representation of the phase diagram of the periodically driven Chern insulator. The red circles (blue squares) depict critical points with gap closings in the quasienergy spectrum at $\omega=0(\pi)$ accompanied by change in $\nu_0 (\nu_\pi)$ shown by the color code which also shows the $\mcl{C}$.}
\label{fig:phasediag_clean}
\end{figure}

We further verify the phase diagram by explicitly calculating $\nu_0$, $\nu_\pi$, and $\mcl{C}$ using Eqs.~\eqref{DisorderFreeInvariant} and \eqref{eq:cn}, and a generic phase diagram for the model is shown graphically in Fig.~\ref{fig:phasediag_clean}.
It is interesting to note that with regard to sequence of $\nu_{0,\pi}$ and $\mcl{C}$ in the phase diagram, there are only two distinct kinds of phase diagrams hosted by the model which correspond to Fig.~\ref{fig:phasediag_clean}(a) and (b). 
While the latter corresponds to the case $4J[\mrm{mod}4\pi]\in[\pi,3\pi]$, the former corresponds to the case otherwise.

\subsection{Non-topological gap closings \label{sec:anomalous_gapclosing}}

\begin{figure}
\includegraphics[width=\columnwidth]{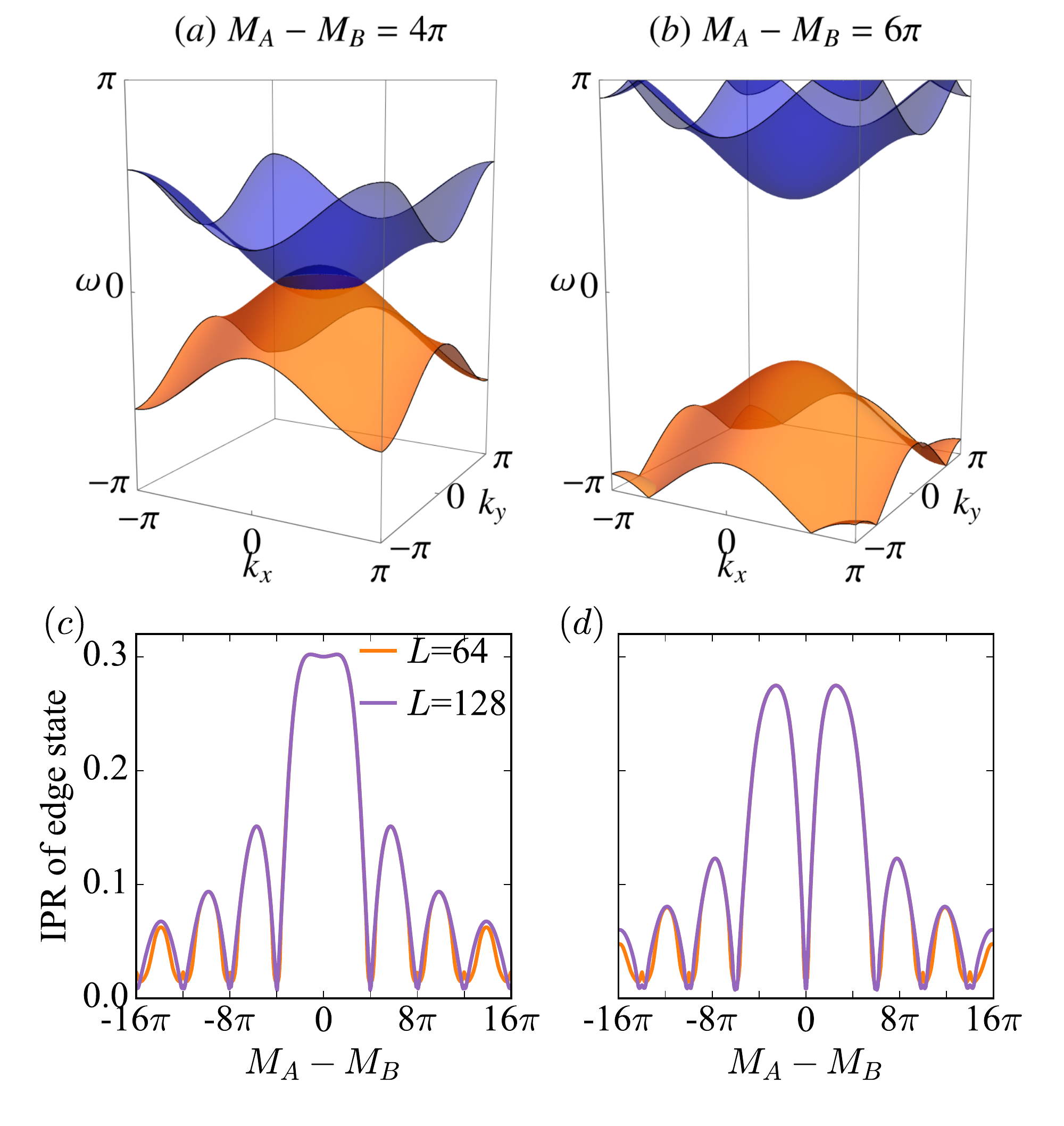}
\caption{Gap-closings in the Floquet quasienergy spectra away from the high-symmetry points at (a) $\omega=0$ and (b) $\omega=\pi$. The IPR of the edge states in the (c) 0-gap and (d) $\pi$-gap on a strip of width $L$, showing the breakdown of edge states at the gap-closings. We use $M_A+M_B = 3J$ for (a) and (c), and $M_A+M_B=2\pi-3J$ for (b) and (d), with $J=\pi/6$.}
\label{fig:ano_gapclosing}
\end{figure}
Having exhaustively studied the topological phase transitions of the Floquet system, we now turn our attention to certain gap-closings of the Floquet quasienergy spectrum which do not lead to any change in the topology of the Floquet bands.
Since these gap-closings happen away from the high-symmetry points, generically we have $\hat{\mathbf{d}}_A\wedge\hat{\mathbf{d}}_B\ne0$, hence its coefficient in Eq.~\eqref{eq:d_eff} has to vanish, implying $\sin d_A=0$ and/or $\sin d_B=0$. 
Inspection of Eq.~\eqref{eq:d_eff} reveals that for $\mathbf{d}_\mrm{eff}$ to vanish, both of them have to vanish simultaneously.
Hence, formally the solutions of these gap-closings can be obtained from the family of solutions of the system of equations
\eq{
	2J^2(1+\cos k_x&\cos k_y)+M_A^2 \nn\\
	&-2JM_A(\cos k_x +\cos k_y) = 4n^2\pi^2,\label{eq:nt1}\\
	2J^2(1+\cos k_x&\cos k_y)+M_B^2 \nn\\
	&-2JM_B(\cos k_x +\cos k_y) = 4m^2\pi^2\label{eq:nt2},
}
where $n$ and $m$ are integers.
We do not find a tractable closed form solutions to Eqs.~\eqref{eq:nt1} and \eqref{eq:nt2}, however a numerical analysis of the quasienergy spectrum shows that these gap-closings happen at
\eq{
	M_A-M_B \approx 2n\pi; ~~n\in\mathds{Z}, \vert n\vert\ge2,
}
where the gap-closing occurs at $\omega=0 (\pi)$ for $n$ being even (odd).
Note that these points depend only on the amplitude of the periodic drive and not the mean.
This is consistent with the observation made in Sec.~\ref{sec:tqpt} that the topological properties of the Floquet bands depend only on the mean and not the amplitude of the drive.

Although the topological properties of the Floquet bands, and hence the presence (absence) and the chiralities of the Floquet edge modes do not change across these gap-closings, they do have important bearings on the robustness of the edge modes. 
The localization length of the edge modes is inversely proportional to the minimum gap around the corresponding quasienergy ($0$ or $\pi$) in the spectrum {\textit{i.e.}}
\eq{
	\xi_0^{-1}\propto \mrm{min}[\omega_{+,\vk}];~~~\xi_\pi^{-1}\propto \pi-\mrm{max}[\omega_{+,\vk}],
}
where $\xi_{0(\pi)}$ is the localization length of the edge modes in the $0 (\pi)$-gap. 
Hence, as the gap in the quasienergy spectrum goes down, the localization length increases.
At the point, where the gap closes, $\xi_{0,(\pi)}$ diverges and the edge state vanishes.
By explicitly calculating the inverse participation ratio of the edge states as a function of $M_A-M_B$, it can be seen that it indeed goes to zero at the gap-closings signaling a breakdown of the edge state through divergence of its localization length.
Representative quasienergy spectra showing such gap-closings is shown in Fig.~\ref{fig:ano_gapclosing} which also shows the vanishing inverse participation ratio (IPR) of the edge states at the gap-closings. The IPR of a state $\vert\psi\rangle$ is defined as $\sum_\pr\vert\psi(\pr)\vert^4$. The IPR of a localized state is inversely proportional to the localization length, where as for a delocalized state it is inversely proportional to the system size. Consequently, in Fig.~\ref{fig:ano_gapclosing}, the minima of the IPR indicate the edge-mode delocalizations that are concurrent with the non-topological gap closings. We emphasize that the chiralities of the edge states do not change on across such gap-closings.

\section{Phase diagram in presence of disorder \label{sec:phasediag_dis}}

In this section, we discuss the effect of quenched disorder on the phase diagram of the periodically driven Chern insulator.
We characterize the phases by calculating the winding invariants, $\nu_{0}$ and $\nu_\pi$, generalized for disordered systems.~\cite{KBRD2010,Titum2016}
Further, by analyzing energy resolved level spacing statistics, we show that the mechanism behind the disorder-induced transitions is what is referred to as ``levitation and annihilation’’,~\cite{OAN2007} extended for Floquet systems.
In this work, we assume an uncorrelated Anderson disorder,
\eq{
	M_{\mathbf{r},\alpha} = M + \delta M_{\mathbf{r},\alpha};~~ M_{\mathbf{r},\beta} = -M + \delta M_{\mathbf{r},\beta},\label{eq:disorder}\\
	\nn \delta M_{\mathbf{r},\alpha},~\delta M_{\mathbf{r},\beta}\in [-W/2,W/2].
} 
Note that, the translation invariant part of $M_\pr$ has the staggered structure on the two sublattices, but the disorder does not, and only the translation invariant is driven periodically in time similar to Eq.~\eqref{eq:Mt}. The realization of the disorder $\delta M$ stays constant with time.

\subsection{Topological invariants with disorder \label{sec:fti_dis}}

Since in the presence of disorder, momentum ceases to be good quantum number, the formulation of the topological invariants need to go beyond the Brillouin zone.
Drawing inspiration from Laughlin`s argument~\cite{L1981} for quantized charge transfer in a system with quantum Hall-like edge states under threading of fluxes, we consider our Hamiltonian \eqref{eq:ham_real} with additional time-independent fluxes $\bm{\theta}=(\theta_x,\theta_y)$ threaded through the lattice. 
For a system with periodic boundary conditions on a torus, the fluxes play the role analogous to that of quasimomenta for the superlattice, for which the unit cell is composed of the $2\times L_x\times L_y$ disordered lattice. 
Hence one can define a family of return maps, $\mcl{U}^{\epsilon}_{\bm{\theta}}$, (for each $\bm\theta$),  similar to Eq.~\eqref{eq:returnmap}, but for the disordered driven system by replacing $U(t)$ by $U_{\bm{\theta}}(t)$ which represents the time-evolution operator in the presence of flux $\bm\theta$.
The winding numbers are then defined as 
\eq{
\nu_{\epsilon} = \int_0^1 dt \int& \frac{d^2\bm{\theta}}{8\pi^2}\bigg[\mrm{Tr}(\mcl{U}^{\epsilon \dagger}\partial_t \mcl{U}^{\epsilon}[\mcl{U}^{\epsilon \dagger}\partial_{\theta_x}\mcl{U}^{\epsilon} ,\mcl{U}^{\epsilon \dagger}\partial_{\theta_y} \mcl{U}^{\epsilon}])\bigg],
\label{eq:nu_dis}
}
where $\mcl{U}^{\epsilon}\equiv \mcl{U}^{\epsilon}_{\bm{\theta}}(t)$.
We also checked that within this framework, we indeed obtain $\nu_{0}-\nu_{\pi} = \mcl{C}$, where $\mcl{C}$ is the Chern number of the disordered system defined as
\eq{
	\mcl{C} = \int\frac{d^2\bm{\theta}}{4\pi} \mrm{Tr}(\msr{P}_{\bm\theta}[\partial_{\theta_x}\msr{P}_{\bm\theta},\partial_{\theta_y}\msr{P}_{\bm\theta}]),
\label{eq:chern_dis}
}
where $\msr{P}_{\bm\theta}$ is the projector onto the eigenstates of the Floquet operator having quasienergy eigenvalues $-\pi<\omega<0$.
Note that, in the numerical implementation of Eqs.~\eqref{eq:nu_dis} and \eqref{eq:chern_dis}, a certain amount of disorder averaging was necessary to wash out the fluctuations due to finite size effects.
However, since the quantities are topological invariants, no disorder averaging is deemed necessary in the thermodynamic limit.~\cite{HM2015}

\subsection{Features of the phase diagram \label{sec:featurespd_dis}}
 \begin{figure}
 \includegraphics[width=\columnwidth]{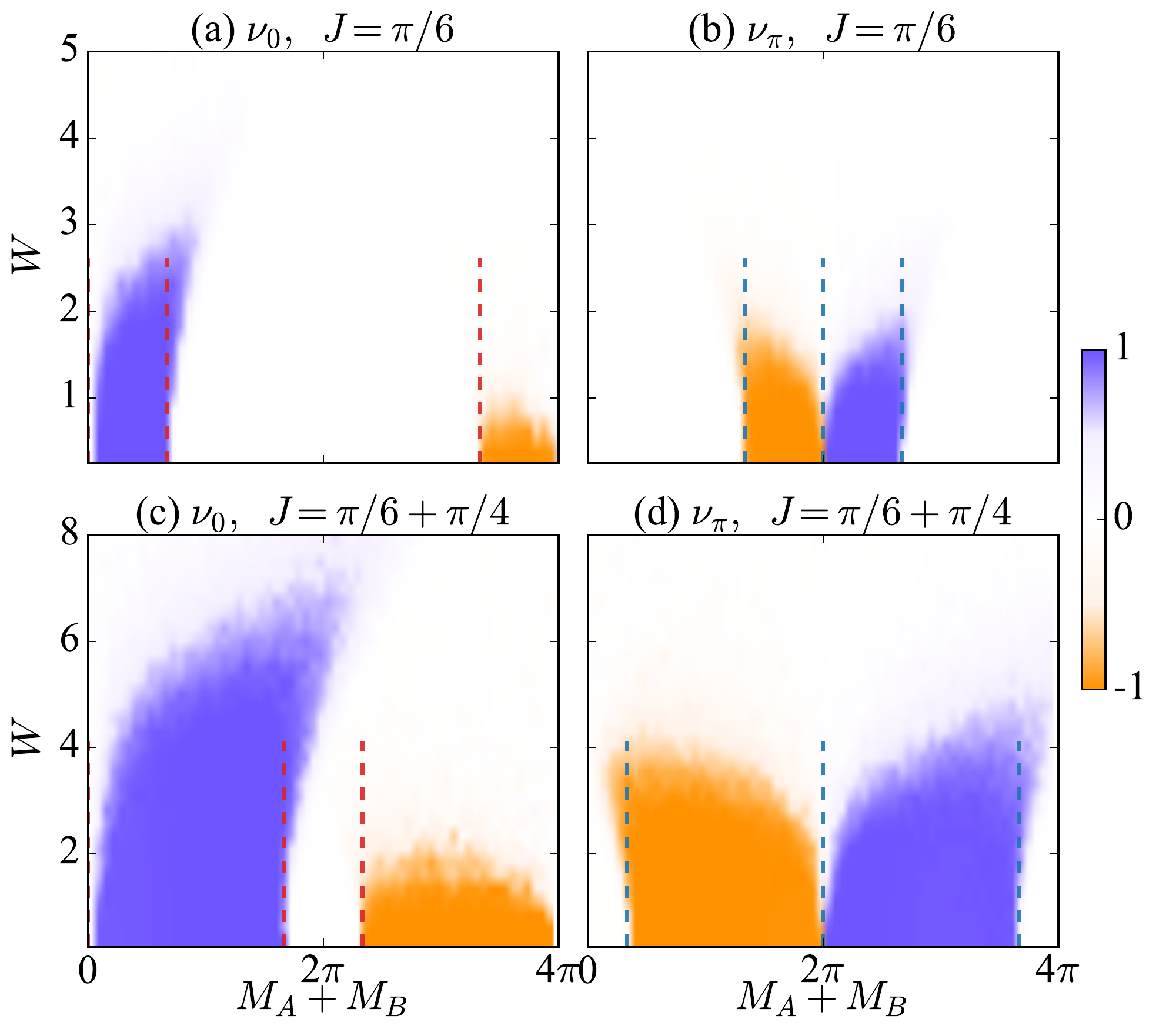}
 \caption{Phase diagram in terms of $\nu_0$ and $\nu_\pi$ are shown in the presence of disorder. The red(blue) dashed lines show the critical points in the absence of disorder. The plots correspond to the $M_A-M_B=4$ line in the $M_A$-$M_B$ plane. For the numerics we used a $2\times 10 \times 10$ system averaged the data over 1000 disorder realizations.}
 \label{fig:phasediag_dis}
 \end{figure}
We consider parameters corresponding to the two qualitative kinds of phase diagrams for the translation invariant system (Fig. \ref{fig:phasediag_clean}(a)-(b)), and study the effect of disorder on the phases by numerically calculating $\nu_{0/\pi}$ using Eq.~\eqref{eq:nu_dis}.
Representative results are shown in Fig.~\ref{fig:phasediag_dis}. Below we discuss the key features of the phase diagram of the system.

Consistent with the general idea of topological invariance and protection of edge states, we find that weak disorder does not affect the phases of the system. However, starting from a topological phase at zero disorder, the system transitions to a trivial one at strong disorder. Analysis of level statistics indicate that the system is fully Anderson localized above the transition.

The critical disorder strength for the transition is lower for systems with parameters $M_{A,B}$ near a phase with an opposite winding number, as compared to systems with parameters close to a trivial phase, indicating that the topological phase in the latter case is much more robust to disorder than the former. This leads to a ``V''-like shape of the phase boundaries, for instance at $M_A+M_B=0$ in Fig.~\ref{fig:phasediag_dis}(a) and (c), and $M_A+M_B=2\pi$ in Fig.~\ref{fig:phasediag_dis}(b) and (d). 
An intuitive explanation for the shape could be obtained from a long-wavelength picture, in which the system at critical disorder strengths can be described as made of a distributions of topological and trivial clusters. Changing the parameter $M_A+M_B$ closer to the opposite topological phase results in introduction of clusters of the opposite winding number. At a coarser scale, these clusters of opposing phases act as a trivial phases increasing the effective density of the trivial clusters. The result is that the transition occurs at a lower disorder strength for systems close to a topological phase with opposite winding number.

For a zero disorder system in the trivial phase, introduction of disorder leads to complete localization of the bulk states. However, for $M_{A,B}$ in a trivial phase but close to the topological phases, the system surprisingly undergoes a transition into a topological phase at intermediate disorder strengths. Such disorder induced topological phases, dubbed as topological Anderson insulators have been previously reported in Floquet systems~\cite{Titum2015} and in various static systems~\cite{jain_tai,PhysRevB.80.165316} and explained via a renormalized mass of the disorder averaged medium~\cite{groth2009theory}.
The Floquet topological Anderson insulator phases appear in continuum with the neighboring topological phases, to the extent that our numerics can resolve. This is unlike the case of static systems with fixed electron densities\cite{groth2009theory}. Such disorder induced phases can occur for systems with both, $\nu_\pi$ and $\nu_0$ topological order.

\begin{figure}
 \includegraphics[width=\columnwidth]{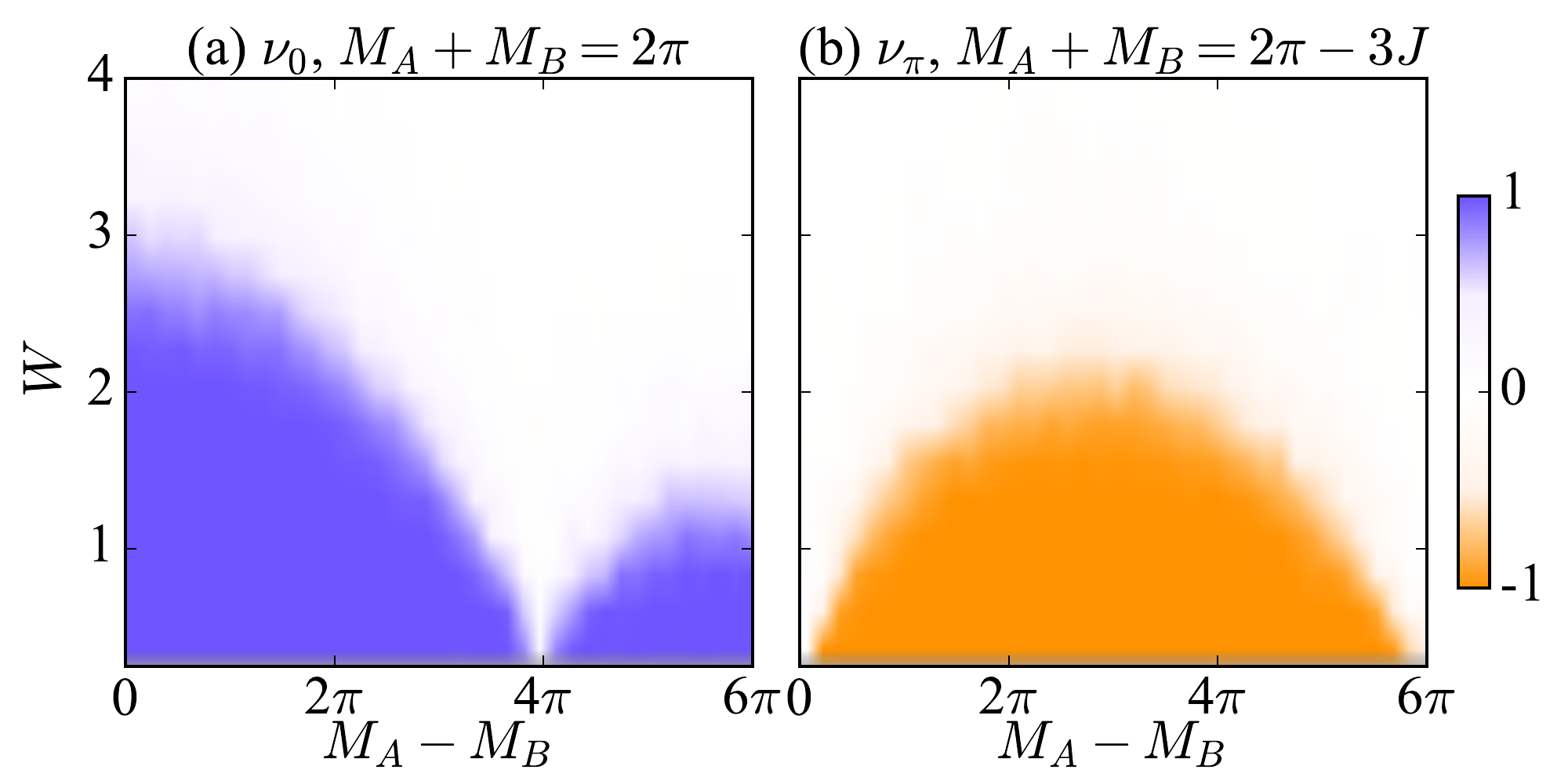}
 \caption{Robustness of edge modes (winding invariants) to disorder. The critical disorder for the transition from the topological to the trivial phase follows the same qualitative behavior as the inverse participation ratio of the zero disorder edge state (Fig.~\ref{fig:ano_gapclosing}) indicating that the robustness of the disorder is intimately connected to the gap in the zero disorder Floquet quasienergy spectrum. The parameters used for the plots are the same as in Fig.~\ref{fig:ano_gapclosing}.}
 \label{fig:ano_gapclosing_dis}
 \end{figure}

Now we discuss the aspects of the phase diagram as a function of $M_A-M_B$. As discussed in Sec.~\ref{sec:anomalous_gapclosing}, there are no topological transitions when $M_A-M_B$ is varied keeping $M_A+M_B$ fixed, however, at certain values of $M_A-M_B$, the gaps close without a topological transition.
We find that this is reflected in the critical disorder strength (for the transitions out of the topological phases). We show this in Fig.~\ref{fig:ano_gapclosing_dis} by using the parameters considered in Fig.~\ref{fig:ano_gapclosing}(c)-(d) and study the effect of disorder on the phase diagram as function of $M_A-M_B$. Such modulations in the critical disorder strength arising from such non-topological gap closings also explain the relative sizes of the lobes of the topological phases shown in Fig. \ref{fig:phasediag_dis}.

Note that the physics arising from interplay of disorder with such non-topological gap closings is qualitatively distinct from that in the case of topological gap closings occurring for example between the topological and trivial phases in Fig.~\ref{fig:phasediag_dis}(red/blue lines). While the topological order is unstable to disorder in the vicinity of the former, the latter is associated with robust topological order till very high disorder as well as formation of a topological Anderson insulator.

\subsection{Level spacing statistics \label{sec:lss}}
%
 \begin{figure}
 \includegraphics[width=\columnwidth]{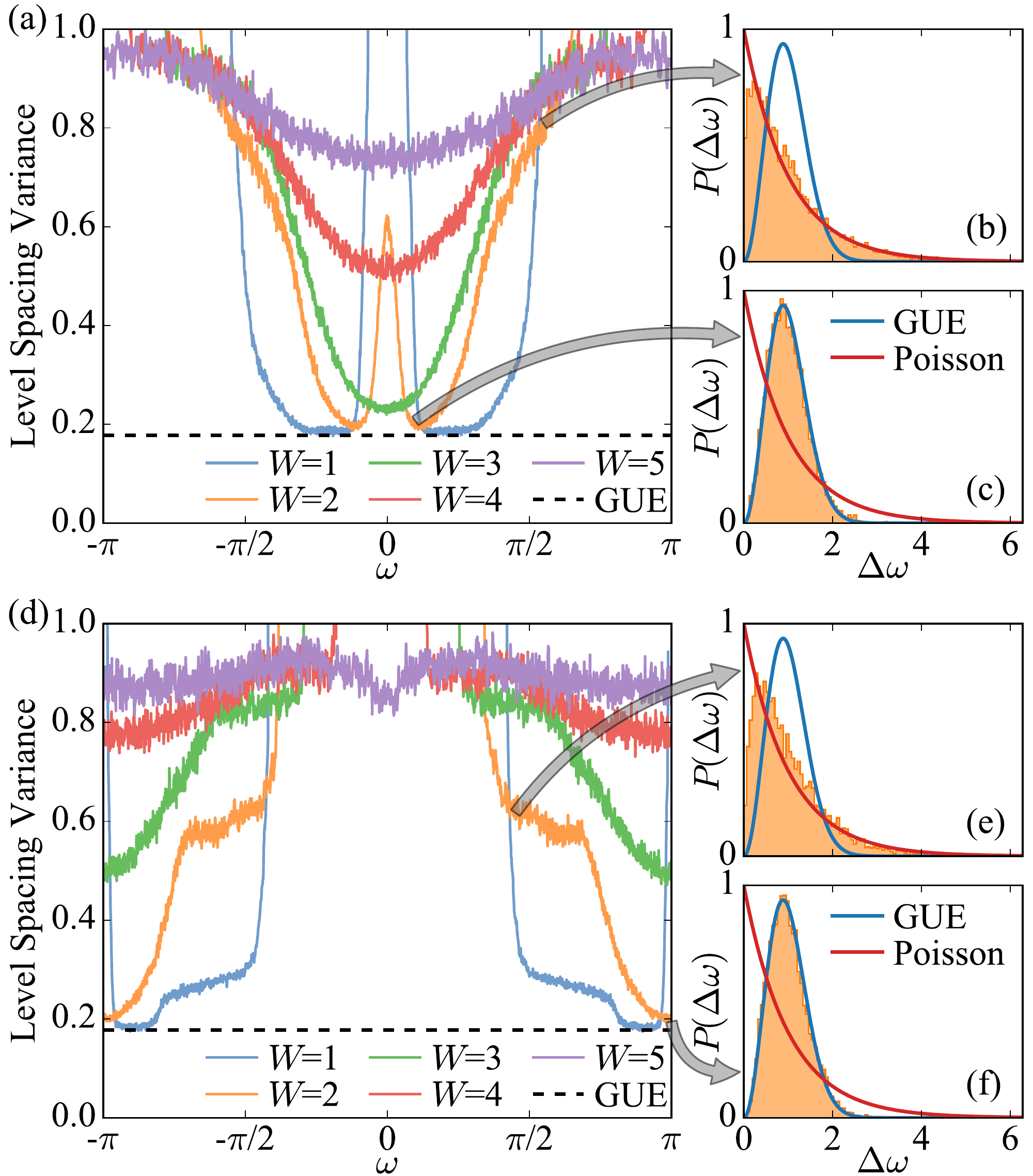}
 \caption{(a) Level spacing statistics for different disorder strengths ($W$) for the case corresponding to (a)$\nu_0=1$ and (d)$\nu_\pi=-1$. As $W$ is increased, the delocalized states indicated by their level spacing variance being that of the GUE, move towards (a)$\omega=0$ and (d)$\omega=\pm\pi$, before finally annihilating and localizing. For $W=2$, $P(\Delta\omega)$ is shown for the localized ((b) and (e)) and delocalized ((c) and (f)) part of the spectrum as indicated by the arrows. For the plots (a)-(c), we use $M_A=-1.5$, $M_B=2.5$ and $J=\pi/6$ where as for (d)-(f) $M_A=0.5$, $M_B=4.5$ and $J=\pi/6+\pi/4$. The numerics are performed on a $2\times 48 \times 48$ sized system and the statistics are taken over 1000 realizations.}
 \label{fig:lss}
 \end{figure}
The interplay of Anderson localization and topology in Chern insulators leads to the occurrence of (at least one) delocalized state(s) in the bulk. Existence of a delocalized state can be motivated from spectral flow arguments.~\cite{L1981,halperin1982} Transitions out of the topological phase, such as the ones induced by disorder is accompanied by a break down of this delocalized state. Except in certain fine tuned scenarios, this happens through what is called a ``levitation and pair annihilation'' mechanism for disorder driven topological phase transition.~\cite{halperin1982, Levine1983, Laughlin1984, OAN2007}

We find essentially the same physics in the Floquet systems that we consider here. The presence of the delocalizes states in the quasienergy spectrum can be inferred from level-spacings as described later in this section. At zero disorder, all single particle states of the system are delocalized. Addition of weak disorder leads to localization of all states in the bulk of the system, leaving a narrow band of delocalized bulk states surrounding every gap that can support edge states. Adding intermediate disorder to a Floquet topological phase leads to formation of bulk localized states in the $0$- as well as the $\pi$- gap of the Floquet quasienergy spectrum.

On increasing the disorder, the delocalized states drift towards each other in the quasi-energy spectrum, and they meet and `annihilate' at the critical disorder strength. In a Floquet system, since the quasienergies are periodic, the delocalized states can in principle levitate along two possible directions, however in all cases that we considered, we found that the delocalized states levitate towards the center of the gap that separates them. The same mechanism appears to apply independently to the gaps around $0$ and $\pi$ (Fig.~\ref{fig:la}).

The quasienergies of the delocalized states can be probed using quasienergy resolved level spacing statistics as was done for the energy spectrum in a static case.~\cite{PHB2010,CLM2015,CGLM2016}
We define the level spacing at quasienergy $\omega$ as $\Delta\omega = \omega_{i+1}-\omega_{i}$, where $\omega_{i}$ is the quasienergy for a finite system closest to $\omega$. These spacings are normalized by disorder averaged level spacings near $\omega$.
If $\omega$ corresponds to a localized part of the spectrum, then the quasienergy values arise from a Poisson process as they are uncorrelated and hence $\Delta\omega$ follows an exponential distribution,\cite{book_M2004} {\textit{i.e.}}, $P(\Delta\omega)\sim e^{-\Delta\omega}$.
On the other hand if the Floquet eigenstates at quasienergy $\omega$ are delocalized, then the quasienergies repel each other and $P(\Delta\omega)$ follows a Wigner-Dyson distribution, specifically a Gaussian Unitary Ensemble\cite{DAlessio2014, book_M2004} (GUE) as the Chern insulators we work with have no symmetries, {\textit{i.e.}}, $P(\Delta\omega)\sim \Delta\omega^2 e^{-4\Delta\omega^2/\pi}$. The two distributions can be distinguished by analyzing the sample variances of the normalized level spacings over many disorder realizations. The delocalized states are indicated by a level spacing variance of $\approx 0.178$ (variance of the Wigner Dyson distribution), whereas fully localized states should show a variance of $1$ (variance of the exponential distribution). However, in our finite system studies, any variance that deviates from the GUE value will be interpreted as indicative of localization.

Representative results of level spacing analysis, that support the levitation annihilation picture are shown in Fig.~\ref{fig:lss}.  On increasing the disorder, the delocalized states move towards each other into the gap, and annihilate each other at the critical disorder.
Although we do not present the results here, the level spacing statistics for the case where both $\nu_0$ and $\nu_\pi$ have finite values, show delocalized states close to both $\omega=0$ and $\omega=\pm\pi$ and localized bulk states away from them, as expected.
%
\begin{figure}
\includegraphics[width=\columnwidth]{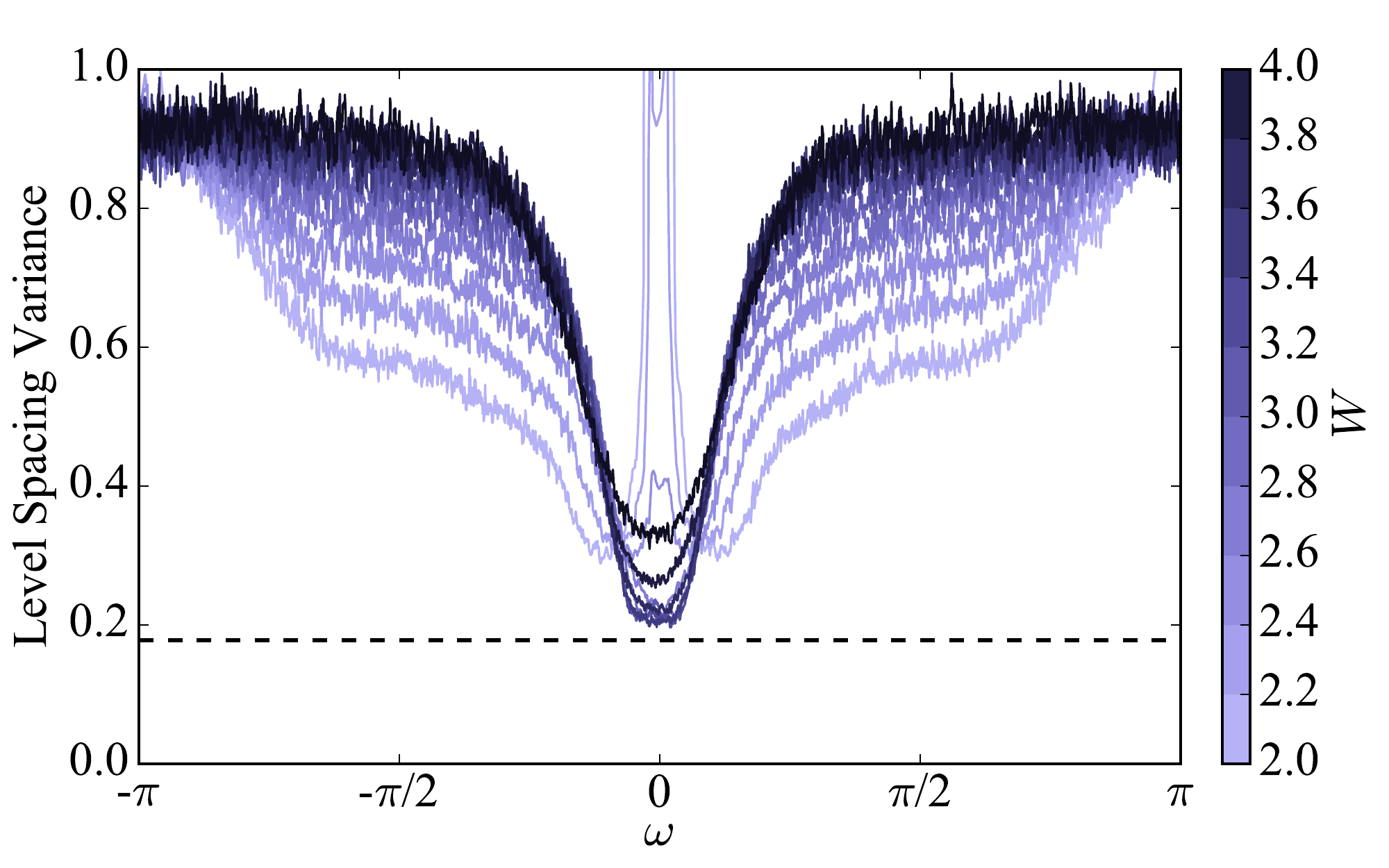}
\caption{Level spacing statistics showing the reentrant topological behavior. The different lines correspond to different disorder strengths (shown by the color bar) and the horizontal dashed value shows the GUE value. At weak disorder there are no delocalized states as the phase is trivial, however on increasing disorder, delocalized states appear close to $\omega=0$ accompanied by a transition to a phase with $\nu_0=1$, and finally at very strong disorder, the system goes to a trivial phase again with all bulk states localized.For the plots we use $M_A=1$, $M_B=2$, and $J=\pi/6$ and the statistics are taken over 1000 disorder realization for a $2\times 48\times 48$ sized system. }
\label{fig:reent_lss}
\end{figure}
%

The level spacing statistics also correctly reflects the reentrant topological behavior. In order to show this we consider parameters corresponding to $M_A+M_B=3$ of Fig.~\ref{fig:phasediag_dis}(a) and analyze the level spacing statistics as shown in Fig.~\ref{fig:reent_lss}. 
It can be seen that at weak disorder, the all the bulk states are localized as there are no quasienergies for which the level spacing variance is close to the GUE value. 
However on increasing the disorder, delocalized states appear which support a finite winding number, before all states localize again at strong disorder.

Hence, the study of level spacing statistics substantiates the physical picture developed in Sec.~\ref{sec:featurespd_dis} and provides evidence for the ``levitation and annihilation'' mechanism of disorder induced phase transitions between Floquet topological and trivial phases. The level spacing studies also adds further evidence supporting the existence of a disorder induced transitions into and out of a topological Anderson insulator phase.

\section{Discussions \label{sec:discussions}}
In summary, we have studied in detail the topological phase diagram of a periodically driven Chern insulator, both in the presence and absence of disorder.
In the absence of disorder, we analytically obtained the exact phase diagram by studying the Floquet quasienergy spectrum and characterized the topology of the phases via appropriate winding invariants.
We found that the topological phase depends only on the mean of the periodic drive and not the amplitude, although the amplitude affects the localization lengths of the chiral edge modes in topological phase. The strict dependence of the phase boundaries (and therefore the phase diagram) on the mean can be understood from the fact that, at those parameters $(\mathbf{k},M_A,M_B,J)$ where the gaps close,
\begin{multline}
U(\mathbf{k},M_A,M_B,J) = e^{\imath H_A} e^{\imath H_B}\propto \mathbb{I}\\
\implies \left[ e^{\imath H_A},e^{\imath H_B} \right]=0
\text{ and } U=e^{\imath (H_A+ H_B)} \propto  \mathbb{I}
\end{multline}
This produces a (sufficient) constraint only on the mean coupling constants. These arguments should hold for a general set of binary drives of coupling constants.

We found that at certain amplitudes there are non-topological gap closings in the quasienergy spectrum leading to vanishing of edge states.
We then extended the phase diagram to include the effects of disorder by numerically computing the winding invariants generalized to include disorder.
The topological phases were found to be robust to weak disorder, however strong disorder induced a phase transition from a topological to trivial phase.
Interestingly, the system also showed a disorder induced transition into a Floquet topological Anderson insulator phase, where the system was trivial at weak/no disorder but underwent a transition to a topological phase at intermediate disorder.
Careful analysis of level spacing statistics of the quasienergy spectrum showed that the disorder-induced transitions happen via a levitation and annihilation of delocalized bulk states within a narrow window of quasienergy in the background of localized bulk states. 

For a topological phase having chiral edge modes with quasienergies in the $0$- and $\pi$-gap, the delocalized bulk states are also at quasienergies close to $0$ and $\pi$ respectively. 
On increasing disorder, the window of delocalized states drifts towards $\omega=0$ and $\omega=\pi$ in the respective case, and at the critical disorder, they meet and annihilate each other driving the system to a trivial phase.
The presence of the delocalized states is necessitated due to the fact that, in the model studied, any topological phase is accompanied by finite Chern number of the bulk bands, though the Chern number does not fully characterize the topological phase.
This is crucially different from the anomalous Floquet-Anderson insulator introduced in Ref.~[\onlinecite{Titum2016}] where all bulk states are localized, hence the bulk bands have zero Chern number with edge modes present at all quasienergies.

There is however an interesting regime in the model studied here where bulk bands with zero Chern number and chiral edge modes coexist.
If the parameters are tuned to a regime, where there exist edge modes in both the gaps at weak disorder, for instance $M_A+M_B =\pi$ in Fig.~\ref{fig:phasediag_dis}(c) and (d), then there are two sets of delocalized bulk states in each band (at weak disorder, the quasienergy spectrum still has two bands), close to $\omega=0$ and $\omega=\pi$.
On increasing disorder, these delocalized states levitate towards their respective gaps.
There is threshold disorder where the gaps and the delocalized states corresponding to one of the edge states (the $\pi$-modes in this case), annihilate while the edge modes in the other gap are still present. 
In such a scenario, the two bands are not well separated, and the bulk states form one continuous band with zero Chern number but with equal number of chiral edge modes on either side, thus realizing a situation similar to the topological anomalous Floquet-Anderson insulator of Ref.~[\onlinecite{Titum2016}].
An important difference though is, unlike Ref.~[\onlinecite{Titum2016}], the system would not realize a quantized charge pump due to the presence of delocalized bulk modes.
On further increasing the disorder, the system goes directly to a trivial Anderson insulator and not an anomolous Floquet topological insulator.

This leads to an important observation that in our case, the delocalized bulk states always annihilate between two bands, leading to break down of all edge states and associated topological order. It is interesting to ask, if there are scenarios where the delocalized bulk states within the same band annihilate each other. The latter situation could potentially lead to a coexistance of fully localized, zero Chern number bulk bands but with chiral edge modes at all quasienergies. A possibility is that the large bandwidth of the bulk bands in our system prevents levitation of the delocalized states through the bulk. This raises an interesting question - namely the fate of the levitation and annihilation mechanism for periodically driven topological systems upon flattening of the Floquet quasienergy bands.

\acknowledgements
We thank R.~Moessner and A.~Lazarides for illuminating discussions and many useful comments in the course of the work.

\appendix
\section{Effective Hamiltonian in the vicinity of high-symmetry points {\label{sec:d_eff}}}
In this Appendix, we present the explicit expressions for the effective Hamiltonian in the vicinity of the high-symmetry points which explicitly shows the sign of $\mrm{Det}[\msr{A}]$ around each high-symmetry point.
In the vicinity of $\vk=(0,0)$, the effective Hamiltonian has the form
\eq{
\sum_{i,j=x,y}\kappa_i\msr{A}_{ij}\sigma^j + \lambda\sigma^z,
}
with 
\eq{
\msr{A}_{ij} = -J & \bigg[\frac{\sin \left(J-\frac{M_A}{2}\right) \cos \left(J-\frac{M_B}{2}\right)}{2 J-M_A}\\
&+\frac{\cos \left(J-\frac{M_A}{2}\right) \sin \left(J-\frac{M_B}{2}\right)}{2 J-M_B}\bigg]\delta_{ij},
}
and $\lambda = \sin[(M_A+M_B-4J)/2]$.
Consequently, $\mrm{Det}[\msr{A}] = 2 A_{xx}^2 >0$.
Similarly, close to $\vk=(\pi,\pi)$ 
\eq{
\msr{A}_{ij} = -J & \bigg[\frac{\sin \left(J+\frac{M_A}{2}\right) \cos \left(J+\frac{M_B}{2}\right)}{2 J+M_A}\\
&+\frac{\cos \left(J+\frac{M_A}{2}\right) \sin \left(J+\frac{M_B}{2}\right)}{2 J+M_B}\bigg]\delta_{ij},
}

and $\lambda = \sin[(M_A+M_B4J)/2]$,  again leading to $\mrm{Det}[\msr{A}] = 2 A_{xx}^2 >0$.
Finally around $\vk=(0,\pi)$,
\eq{\msr{A}_{ij} = -J&\bigg[\frac{\cos(M_B/2)\sin(M_A/2)}{M_A} \nn\\ &+\frac{\cos(M_A/2)\sin(M_B/2)}{M_A }\bigg]\delta_{ij}(\delta_{ix}-\delta_{iy})
}
and $\lambda = \sin[(M_A+M_B)/2]$. Consequently $\mrm{Det}[\msr{A}] = -2 A_{xx}^2 <0$.

\bibliography{references}

\end{document}